\documentclass{article}

\usepackage{arxiv}

\usepackage[utf8]{inputenc} 
\usepackage[T1]{fontenc}    
\usepackage{hyperref}       
\usepackage{url}            
\usepackage{booktabs}       
\usepackage{amsfonts}       
\usepackage{nicefrac}       
\usepackage{microtype}      
\usepackage{lipsum}		    
\usepackage{graphicx}
\usepackage{natbib}
\usepackage{doi}
\usepackage{siunitx}

\title{Ultra-stable transportable ultraviolet clock laser using cancellation between photo-thermal and photo-birefringence noise}

\author{ \href{https://orcid.org/0000-0003-4995-086X}{\includegraphics[scale=0.06]{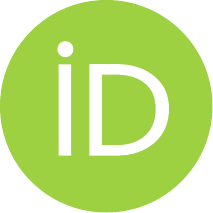}\hspace{1mm}Benjamin Kraus}\\
	Physikalisch-Technische Bundesanstalt\\
    Bundesallee 100\\
	38116 Braunschweig \\
    Germany\\
    \And
    \href{https://orcid.org/0000-0001-8522-4457}{\includegraphics[scale=0.06]{orcid.pdf}\hspace{1mm}Sofia Herbers}\\
	Physikalisch-Technische Bundesanstalt\\
    Bundesallee 100\\
	38116 Braunschweig \\
    Germany\\
    \And
    \href{https://orcid.org/0000-0002-9645-3818}{\includegraphics[scale=0.06]{orcid.pdf}\hspace{1mm}Constantin Nauk}\\
	Physikalisch-Technische Bundesanstalt\\
    Bundesallee 100\\
	38116 Braunschweig \\
    Germany\\
    \And
    \href{https://orcid.org/0000-0001-5661-769X}{\includegraphics[scale=0.06]{orcid.pdf}\hspace{1mm}Uwe Sterr}\\
	Physikalisch-Technische Bundesanstalt\\
    Bundesallee 100\\
	38116 Braunschweig \\
    Germany\\
    \And
    \href{https://orcid.org/0000-0002-4705-8854}{\includegraphics[scale=0.06]{orcid.pdf}\hspace{1mm}Christian Lisdat}\\
    Physikalisch-Technische Bundesanstalt\\
    Bundesallee 100\\
    38116 Braunschweig \\
    Germany\\
	\And
	\href{https://orcid.org/0000-0003-0773-5889}{\includegraphics[scale=0.06]{orcid.pdf}\hspace{1mm}Piet O. Schmidt} \\
	Physikalisch-Technische Bundesanstalt\\
    Bundesallee 100\\
	38116 Braunschweig \\
    Germany\\
    Institute of Quantum Optics\\
    Leibniz University Hannover\\
    30167 Hannover\\
    Germany\\
	\texttt{ Piet.Schmidt@quantummetrology.de} \\
}

\renewcommand{\undertitle}{\textcopyright\, 2025 Optica Publishing Group. Users may use, reuse, and build upon the article, or use the article for text or data mining, so long as such uses are for non-commercial purposes and appropriate attribution is maintained. All other rights are reserved. \href{https://doi.org/10.1364/OL.544907}{https://doi.org/10.1364/OL.544907}}
\renewcommand{\shorttitle}{\textit{arXiv} Template}

\hypersetup{
pdftitle={A template for the arxiv style},
pdfsubject={q-bio.NC, q-bio.QM},
pdfauthor={David S.~Hippocampus, Elias D.~Striatum},
pdfkeywords={First keyword, Second keyword, More},
}

\begin{document}
\maketitle

\begin{abstract}
Optical clocks require an ultra-stable laser to probe and precisely measure the frequency of the narrow-linewidth clock transition.
We introduce a portable ultraviolet (UV) laser system for use in an aluminum quantum logic clock, demonstrating a fractional frequency instability of approximately $\mathrm{mod}\,\sigma_\mathrm{y} = 2 \times 10^{-16}$. The system is based on an ultra-stable cavity with crystalline AlGaAs/GaAs mirror coatings, alongside with a frequency quadrupling system employing two single-pass second harmonic generation (SHG) stages. Its acceleration sensitivity, measured in all three axes, does not exceed $4(2) \times 10^{-12}$/(ms$^{-2}$) and is among the lowest recorded for transportable systems to date.  
Additionally, partial cancellation between photo-thermal noise and photo-birefringence noise is used to effectively mitigate noise induced by intra-cavity optical power fluctuation at lower Fourier frequencies.
\end{abstract}


Optical clocks have achieved systematic fractional frequency uncertainties on the order of $1\times10^{-18}$ \cite{sanner_optical_2019,bothwell_jila_2019,mcgrew_atomic_2018,ohmae_transportable_2021,zhiqiang_176lu_2023
}, with the aluminum quantum logic clock \cite{brewer_27+_2019} among the most accurate ion clocks. This remarkable advancement in accuracy calls for a redefinition of the SI unit second \cite{dimarcq_roadmap_2024} and extends applications in tests of fundamental physics \cite{sanner_optical_2019,safronova_search_2018,roberts_search_2020,filzinger_improved_2023}. Transportable clocks can be used for chronometric leveling in relativistic geodesy  and monitoring of geodynamic processes \cite{grotti_long-distance_2024, chou_optical_2010, mcgrew_atomic_2018, yuan_demonstration_2024, mehlstaubler_atomic_2018, giuliani_determination_2024}. The stability of single ion clocks is typically fundamentally limited by quantum projection noise (QPN) \cite{itano_quantum_1993}. With typical interrogation times of \SI{100}{\milli\s}, this results in a stability on the order of $10^{-15}/\sqrt{\tau/s}$ \cite{brewer_27+_2019, peik_laser_2006}, requiring weeks of measurement time to reach $1\times10^{-18}$ levels of statistical uncertainty corresponding to \SI{1}{\centi\meter} height resolution in relativistic geodesy. 
Smaller instabilities can be achieved by longer interrogation times \cite{leroux_-line_2017}, enabled by a clock laser with sufficiently long coherence time.

Here, we introduce a transportable UV laser system \cite{kraus_highly_2023} designed for the use in a transportable $^{27}\mathrm{Al}^+$ quantum logic clock. The system includes an ultra-stable cavity with crystalline Al$_{0.92}$Ga$_{0.08}$As/GaAs mirror coatings \cite{cole_tenfold_2013} on fused silica (FS) mirror substrates with ultra-low expansion (ULE\textsuperscript{\textregistered}) glass compensation rings \cite{leg10}. The mirrors are separated by a ULE\textsuperscript{\textregistered} glass spacer of \SI{20}{\centi\meter} length, and the mounting structure is based on the design by Herbers et al. \cite{herbers_transportable_2022}. The cavity's expected thermal noise limit (including Brownian noise, thermo-elastic noise and thermo-refractive noise) is approximately $\mathrm{mod}\,\sigma_\mathrm{y} = 7 \times 10^{-17}$ at one second, its finesse is $2.02(5) \times 10^5$ and its drift rate is approximatly \SI{30}{mHz\per\s}. 

A seed laser at \SI{1070}{\nano\m} (Koheras ADJUSTIC Y10 from NKT) is stabilized with the Pound-Drever-Hall (PDH) technique to a resonance frequency of the cavity. A frequency quadrupling system, featuring two single-pass SHG stages \cite{kraus_phase-stabilized_2022}, generates UV light at \SI{267.4}{\nano\m} while maintaining the seed light's frequency stability. The IR light distributed to the cavity, the optical comb, and the frequency quadrupling system is phase stabilized on a compact optical breadboard.

\begin{figure}[b]
\begin{minipage}{0.49\textwidth}
     \centering
    \includegraphics[width=0.85\textwidth]{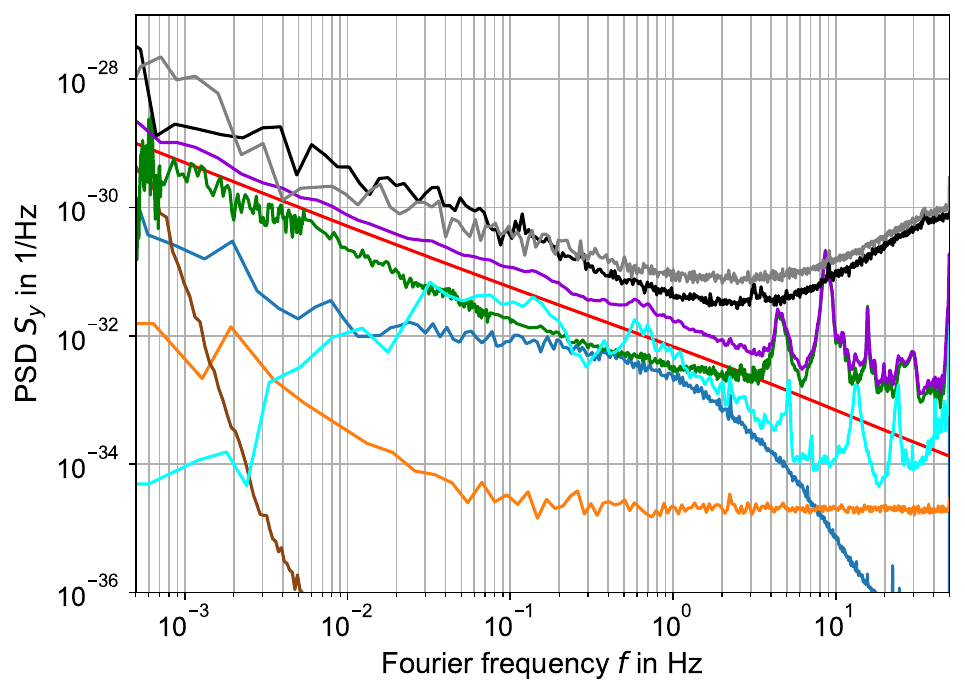}
 \end{minipage}
  \begin{minipage}{0.5\textwidth}
     \centering
  \includegraphics[width=0.85\textwidth]{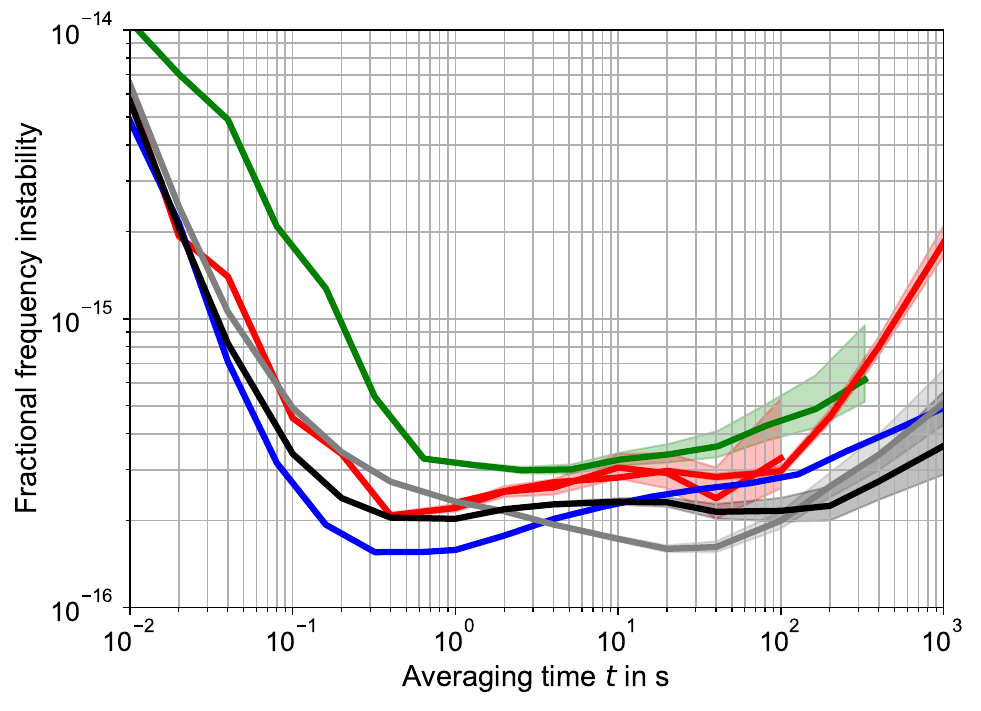} 
  \end{minipage}
\caption{Left: Power spectral density (PSD) of fractional frequency noise of the laser system (black and grey) and the expected frequency instability (violet), encompassing the instability induced by thermal noise (red), seismic noise (cyan), photo-thermal and photo-birefringence noise (blue), RAM (orange), temperature fluctuations (brown) and the reference laser system \cite{matei_15_2017} (green). Right: Allan deviations of the fractional frequency instability of different transportable cavity designs from Herbers et al. \cite{herbers_transportable_2022} (blue), Häfner et al. \cite{hafner_transportable_2020} (green), Xiao et al. \cite{xiao_transportable_2022} (red), and this paper (black and grey). The data sets of the curves from \cite{herbers_transportable_2022, hafner_transportable_2020,xiao_transportable_2022} were provided by the authors.}
\label{fig:lasernoise}
\end{figure}

The characterization of the noise performance of the cavity follows standard procedures in the field as, e.g., described in references \cite{herbers_transportable_2022, yu_excess_2022}.
To characterize the frequency stability of the IR laser system, it was compared to an optical frequency comb, stabilized to a second ultra-stable laser system, which has a fractional frequency instability of approximately $\mathrm{mod}\,\sigma_\mathrm{y} = 4 \times 10^{-17}$ at \SI{1}{\s} \cite{matei_15_2017}. The instability induced by residual amplitude modulation (RAM) \cite{Wong_RAM_1985} with a temperature-stabilized, fiber-coupled waveguide EOM (NIR-MPX-LN-0.1-00 from IXblue) is on the order of $5\times10^{-18}$ at \SI{1}{\s} without an active RAM stabilization scheme. It is below the thermal noise limit of $7\times10^{-17}$ for averaging times between \SI{0.01}{\s} and \SI{100}{\s}. An active RAM stabilization scheme \cite{zhang_reduction_2014} was implemented and RAM-induced instability was further reduced below $6\times10^{-18}$ for averaging times between \SI{0.2}{\s} and \SI{100}{\s} (Figure \ref{fig:lasernoise} top, orange line).

\begin{figure}[b]
\centering
\includegraphics[width=0.5\linewidth]{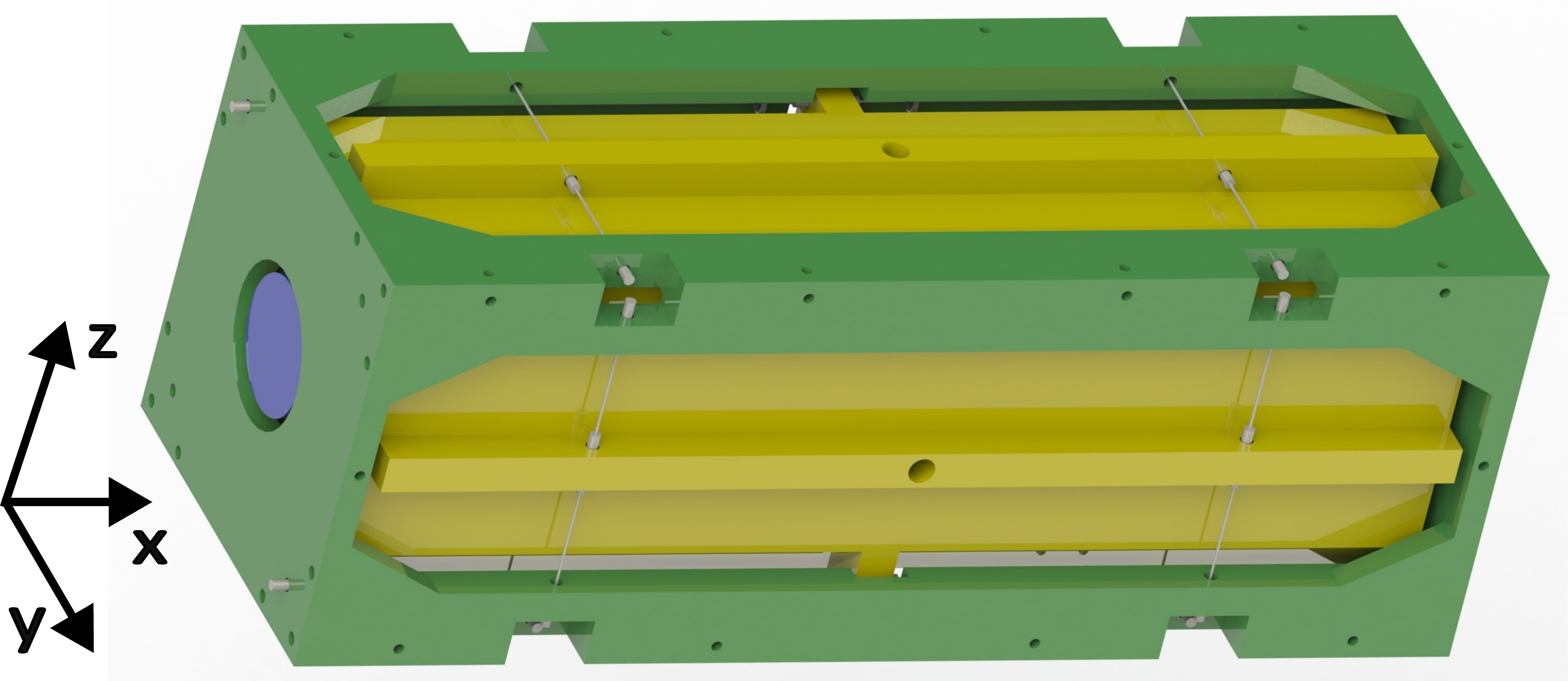}
\caption{3D-model of the cavity, including the mirrors (blue), the ULE spacer (yellow), and the suspension attached to the mounting frame (green).}
\label{fig:cavity}
\end{figure}

\begin{figure*}[ht]
\centering
\includegraphics[width=5.2cm]{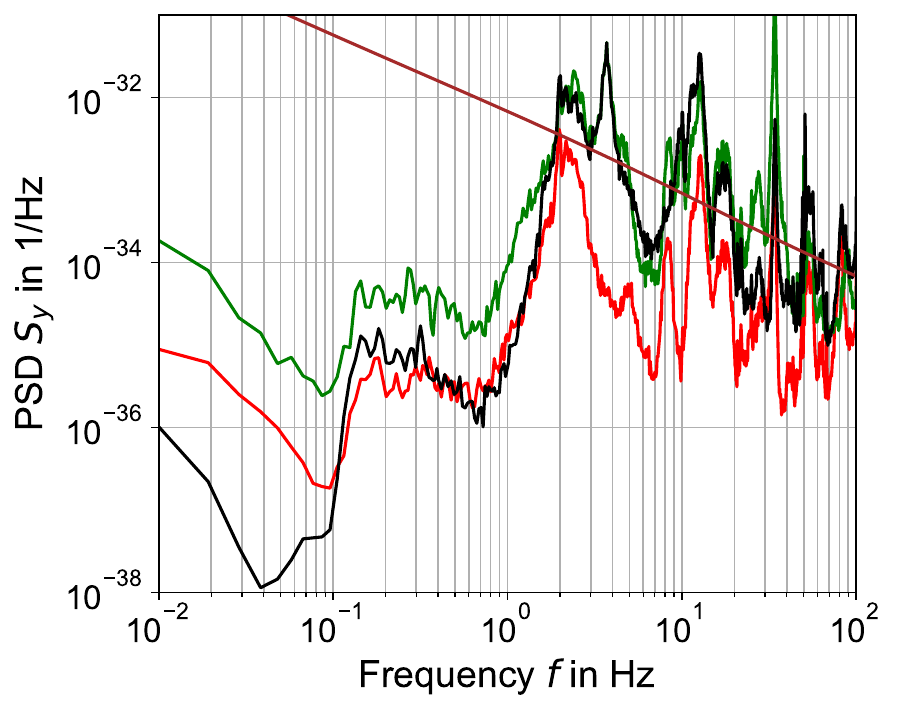}
\includegraphics[width=5.2cm]{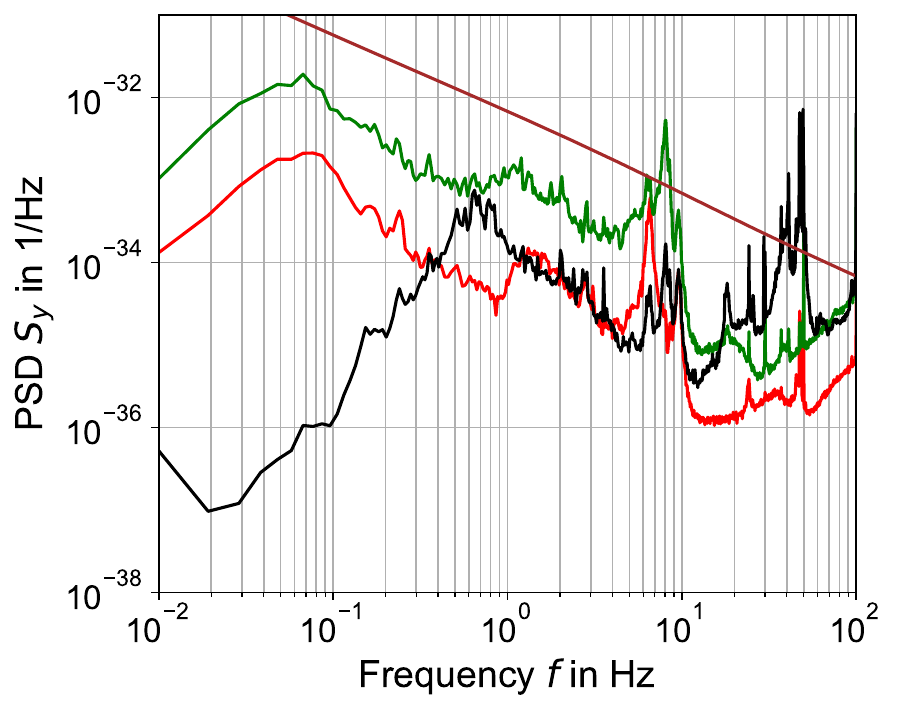}
\includegraphics[width=5.2cm]{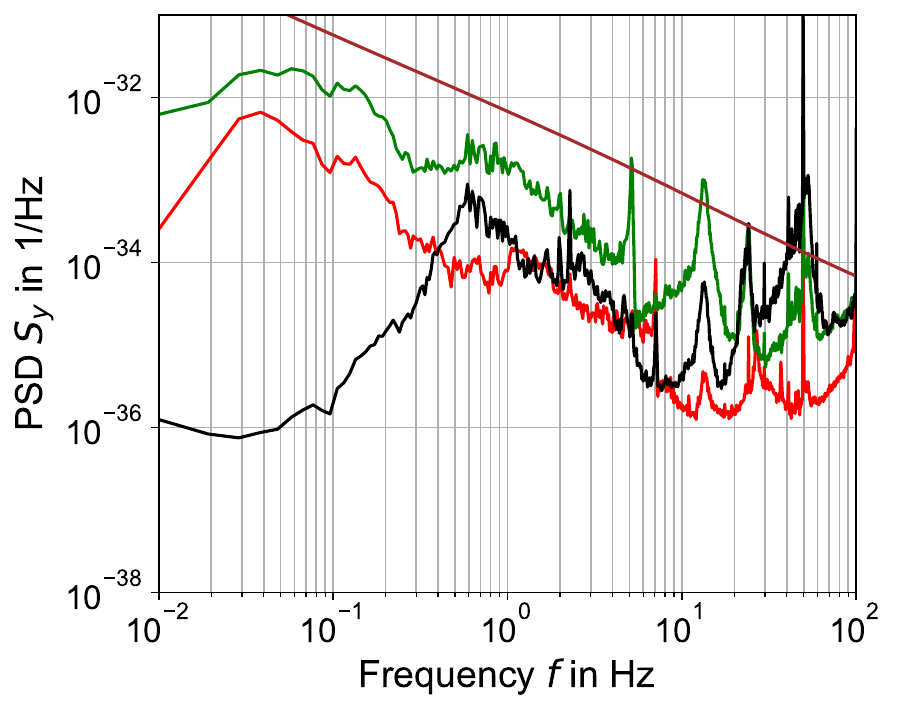}
\caption{Calculated power spectral density (PSD) $S_\text{y}$ of fractional frequency instability induced by mechanical vibration, while the cavity is placed on a passive vibration isolation platform (left), on an active platform (middle), and on the active platform placed inside the rack (right). Legend: Vibration-induced frequency noise along the optical axis (red), horizontally and perpendicular to the optical axis (green) and vertical (black). For comparison the thermal noise limit is shown (brow). }
\label{fig:vibrationnoise}
\end{figure*}

Vibrations impacting the optical length of the cavity can significantly undermine the system's stability. Thus, minimizing the acceleration sensitivity of the cavity is essential. 
The influence of vibrations on the frequency stability of the cavity is reduced by a special mounting structure \cite{uwe_frequenzstabilisierungsvorrichtung_2012, herbers_transportable_2022} (Fig.~\ref{fig:cavity}) and by placing the setup on a commercial vibration isolation platform (see Fig.~S1 in the supplement). 
For the \SI{20}{\centi\m} long horizontally aligned cavity, the measured acceleration sensitivities are $1(1)\times10^{-12}$/(ms$^{-2}$) along the optical axis, $4(2)\times10^{-12}$/(ms$^{-2}$) horizontally and perpendicular to the optical axis, and $3(2)\times10^{-12}$/(ms$^{-2}$) in the vertical direction. These sensitivities are among the lowest in the literature (see Table S1 in the supplement) and agree with those of a similar design \cite{herbers_transportable_2022}.

The frequency noise induced by vibrations was evaluated on both a passive (CT-1 from Minus K) and an active (TS-150 from Table Stable) vibration isolation platform. The passive platform (Fig.~\ref{fig:vibrationnoise} left) exhibits superior performance at lower vibration frequencies, whereas the active platform (Fig.~\ref{fig:vibrationnoise} middle and right) was more effective at higher frequencies. Vibration-induced frequency noise on the active platform exceeds the thermal noise limit at some Fourier frequencies higher than \SI{8}{\hertz}. Using the passive table, the instability caused by vibration exceeds the thermal noise floor already at frequencies around \SI{1}{\hertz} and above. On the active table, no significant increase in vibration-induced frequency noise was measured after the laser system was integrated into a standard 19-inch rack (Fig.~\ref{fig:vibrationnoise} right).

The cavity setup includes a temperature stabilization system, featuring two inner passive heat shields, an active heat shield, an actively temperature stabilized vacuum chamber, and an outer insulation layer. Thermalization time constants of $\tau_1=\SI{47.2}{\hour}$ and $\tau_2=\SI{0.9}{\hour}$ were measured. The temperature of the active heat shield is stabilized to the measured zero-crossing temperature of the cavity at \SI{299.3(2)}{\kelvin}, effectively suppressing cavity length expansion due to temperature fluctuations below the thermal noise limit (Fig.~\ref{fig:lasernoise}).

\begin{figure}[b]
\centering
\includegraphics[width=0.5\linewidth]{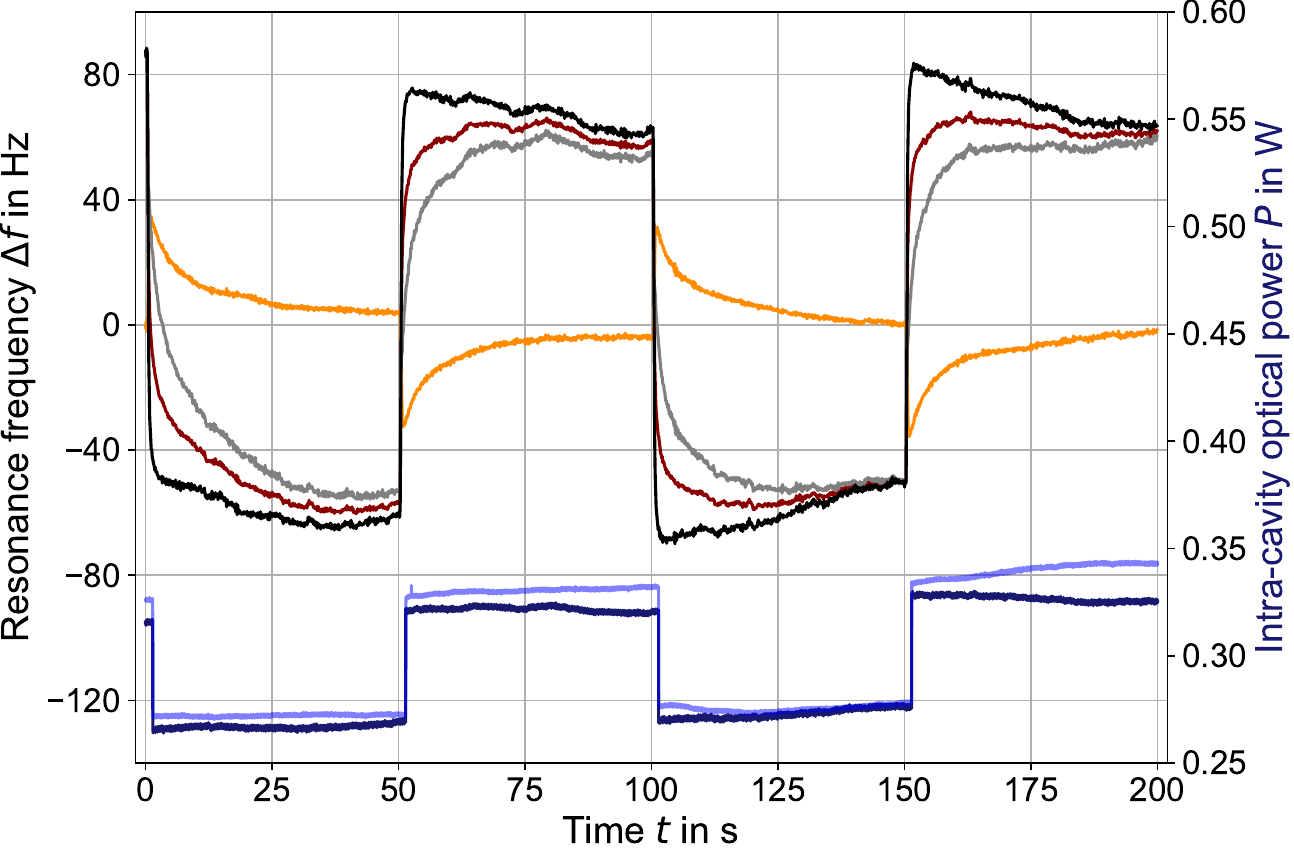}
\caption{Response of the resonance frequency of the cavity to a step function of the optical intra-cavity power for the fast polarization axis (brown) and for the slow polarization axis (orange). Summing the frequency response of both polarization modes eliminates the birefringence effect and represents the photo-thermal effects (gray). Subtracting the frequency response for the slow axis from the frequency response for the fast axis shows the birefringence effect on the polarization line splitting (black). The birefringence effect shows a faster response to power changes than the photo-thermal effects. The optical power modulation is shown for the fast polarization axis (dark blue) and for the slow polarization axis (light blue).}
\label{fig:photojup}
\end{figure}

The birefringence of the crystalline coatings results in a frequency splitting of the polarization modes. In our cavity the mirrors are aligned to maximize the frequency splitting to \SI{345}{\kilo\hertz}. Fluctuations in intra-cavity optical power lead to an optical length change of the cavity via the photo-thermal effect and the photo-birefringence effect. While the photo-thermal effect is experimentally and theoretically well understood \cite{Gorodetsky_Thermalnoise_2008,Farsi_photothermal_2012,Cerdonio_thermoelastic_2001}, the photo-birefringence effect has been experimentally observed in several cavities, but a complete theoretical explanation remains elusive \cite{yu_excess_2022,Kedar:23, ma_ultrastable_2024,Zhu_23}. The  photo-birefringence effect has a negative or positive sign depending on the polarization of the light being aligned with the slow or fast axis of the resonator, respectively. For the cavity presented, the photo-thermal effect encompasses the photo-thermoelastic effect from the substrate and both the photo-thermoelastic effect and the photo-thermorefractive effect from the coating.

Measurements of the photo-induced frequency response were conducted, incorporating both photo-thermal and photo-birefringence effects. The response of the resonance frequency to step changes of the intra-cavity optical power for both the slow, and fast axis was measured (Fig.~\ref{fig:photojup}). The response for light polarized along the fast axis is larger compared to light polarized along the slow axis and the two polarization directions exhibit a different sign as the photo-birefringence effect is affected by the polarisation of the light. Furthermore, the slow-axis response returns to its initial value after several seconds. Similar observations for room temperature cavities and more detailed measurements on this behavior can be found in reference \cite{ma_ultrastable_2024}.

\begin{figure}[hb!]
\centering
\includegraphics[width=0.5\linewidth]{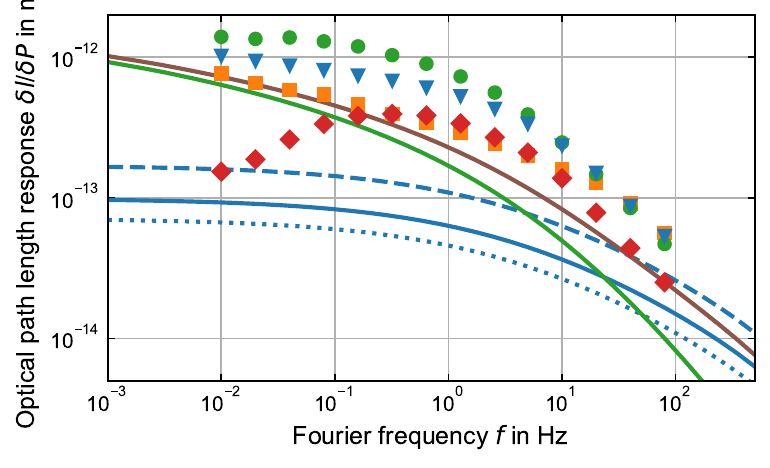}
\caption{Modulus of the response of the optical length of the cavity to a sinusoidal modulation of the optical intra-cavity power for the fast polarization axis with mean optical intra-cavity power of \SI{0.4}{\watt} (green), \SI{0.7}{\watt} (blue) and \SI{1.8}{\watt} (orange) and for slow polarization axis with mean optical intra-cavity power of \SI{0.4}{\watt} (red). The magnitude of the transfer function is shown for the total photo-thermal noise (brown), which is the sum of the photo-thermoelastic contribution of the substrate (green) and the photo-thermal contribution of the coating (solid blue) \cite{Farsi_photothermal_2012}, assuming an absorption coefficient of 2\,ppm for each mirror. The photo-thermal contribution of the coating comprises the photo-thermorefractive coating contribution (dotted blue) and the photo-thermoelastic coating contribution (dashed blue), which have opposite signs.}
\label{fig:photofreq}
\end{figure}

\begin{figure}[hb]
\centering
\includegraphics[width=0.5\linewidth]{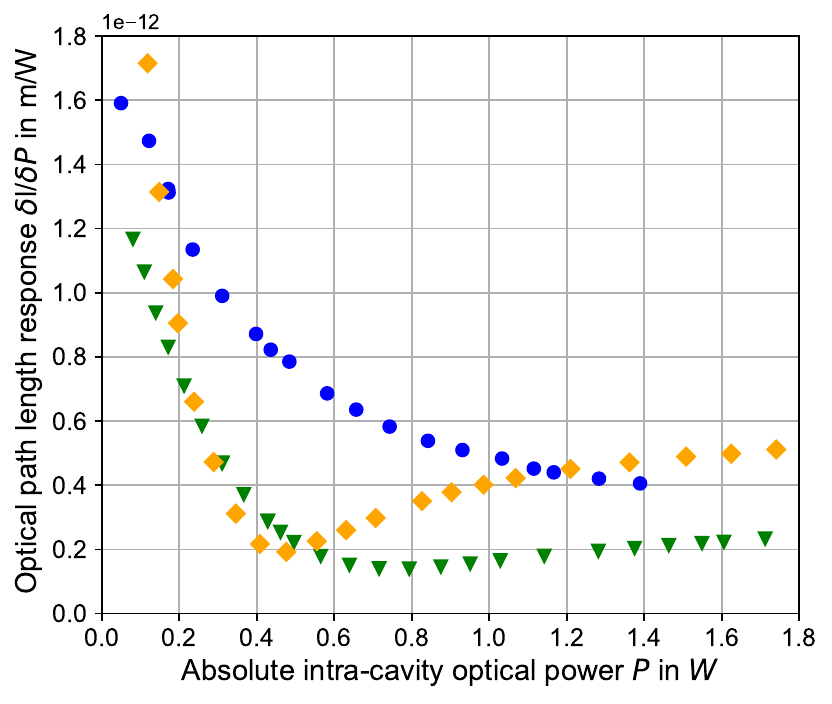}
\caption{Response of the optical length of the cavity to a sinusoidal modulation of the optical intra-cavity power for the fast polarization axis with a modulation frequency of \SI{0.5}{\hertz} (blue), for the slow polarization axis with a modulation frequency of \SI{0.5}{\hertz} (green) and with a modulation frequency of \SI{0.05}{\hertz} (orange).}
\label{fig:photopower}
\end{figure}

Additionally, the photo-induced frequency response was measured for different settings of light polarization and intra-cavity optical power while applying sinusoidal intra-cavity optical power modulation (Fig.~\ref{fig:photofreq} and Fig.~\ref{fig:photopower}). For light polarization aligned to the fast axis, the photo-induced frequency response exceeds the theoretical expectations of the photo-thermal effect as both the photo-thermal effect and the photo-birefringence effect have positive sign. The photo-induced frequency response decreases for higher absolute intra-cavity optical power. For light with polarization aligned to the slow axis, a partial cancellation between the photo-thermal effect and the photo-birefringence effect is achieved, as the photo-thermal effect has a negative sign. Furthermore, the cancellation is maximized by tuning the absolute intra-cavity optical power, as the amplitude of the photo-birefringence effect depends on the absolute power, while the photo-thermal effect depends only on the magnitude of the power change. As the timescales of the two effects are different, the cancellation also depends on the considered Fourier frequency. For a Fourier frequency of \SI{0.5}{\hertz} a minimum of the photo-induced frequency variation is given at \SI{0.8}{\watt} absolute intra-cavity optical power, while a minimum is found around \SI{0.5}{\watt} for a Fourier frequency of \SI{0.05}{\hertz}.     

Suppressing photo-induced frequency noise is most crucial at lower Fourier frequencies. Therefore, the intra-cavity optical power is stabilized to \SI{0.4}{\watt}, at which the optical power fluctuations result in a fractional frequency instability of $4\times10^{-17}$ at \SI{1}{\s}. The photo-induced frequency instability is below the thermal noise limit for Fourier frequencies between \SI{5e-4}{\hertz} and \SI{50}{\hertz} (Fig. \ref{fig:lasernoise} top).

The fractional frequency instability of the laser is approximately $\mathrm{mod}\,\sigma_\mathrm{y} = 2 \times 10^{-16}$ measured on different days, thus emphasizing the reproducibility of the measurements (Fig. \ref{fig:lasernoise} black and grey). The observed frequency noise slightly exceeds the accumulated frequency noise from all discussed noise sources. At Fourier frequencies above a few hertz, the increase of laser frequency instability is caused by limitations in the stability transfer to the reference laser including optical path length stabilization of fibers and an optical frequency comb. 

For frequency conversion from \SI{1069.6}{\nano\m} to \SI{267.4}{\nano\m}, two single-pass SHG stages with an integrated optical path length stabilization scheme were used \cite{kraus_phase-stabilized_2022}. An UV output power of up to \SI{60}{\micro\watt} was reached. With a fractional frequency instability below $10^{-16}$ after an averaging time of more than \SI{1}{\s}, the frequency quadrupling system does not limit the frequency stability of the laser system. Furthermore, the instability of the quadrupling system reaches $10^{-18}$ after \SI{1000}{\s} with an accuracy below $4\times10^{-19}$ and thus does not limit the performance of the optical clock.

In conclusion, a transportable UV laser system for a $^{27}\mathrm{Al}^+$ quantum logic clock was introduced, achieving a fractional frequency instability of $2\times10^{-16}$, on par with the most stable existing transportable laser systems to date \cite{herbers_transportable_2022,hafner_transportable_2020,xiao_transportable_2022} (Fig.~\ref{fig:lasernoise} bottom). The primary frequency noise sources for the cavity were determined, including thermal noise, RAM, vibration noise, temperature fluctuations, photo-thermal noise, and photo-birefringence noise. By tuning the laser light power and polarization, a cancellation of the photo-thermal effect and the photo-birefringence effect was achieved at lower Fourier frequencies, suppressing photo-induced frequency noise below the thermal noise limit. The evaluated frequency instability of the IR laser system revealed additional, unexpected frequency noise, requiring further investigation into other potential sources, particularly birefringence noise, which was recently identified in AlGaAs mirror coatings at cryogenic temperatures \cite{yu_excess_2022,Kedar:23}. 

\textbf{Funding} Deutsche Forschungsgemeinschaft (project-ID 434617780, project-ID 27420014); European Partnership on Metrology (22IEM01, 23FUN03); State of Lower Saxony, Hannover, Germany, through Niedersächsisches Vorab.

\textbf{Acknowledgments}Funded by the Deutsche Forschungsgemeinschaft (DFG, German Research Foundation) under Germany’s Excellence Strategy-EXC-2123 QuantumFrontiers-390837967, SFB 1464 (project-ID 434617780), and SFB 1227 (project-ID 274200144). This joint research project (QVLS-Q1) was financially supported by the State of Lower Saxony, Hannover, Germany through Niedersächsisches Vorab. The projects TOCK (22IEM01) and HIOC (23FUN03) have received funding from the European Partnership on Metrology, co-financed from the European Union’s Horizon Europe Research and Innovation Programme and by the Participating States. This work was supported by the Max Planck, RIKEN, PTB Center for Time, Constants, and Fundamental Symmetries (C-TCFS). We thank Fabian Wolf for the helpful comments on the manuscript, Fabian Dawel for his support in the laboratory, Chun Yu Ma for the discussion of the experimental results, and David Weber, Mattias Missera, Andreas Koczwara, and Andr\'{e} Uhde for building the electronic and mechanical parts of the experiment.

\textbf{Disclosures} The authors declare no conflicts of interest.

\textbf{Data availability} Data underlying the results presented in this paper are available in Ref.~\cite{datax1}.

\textbf{Supplemental document} See Supplement 1 for supporting content.

\bibliographystyle{unsrtnat}






\end{document}